\documentclass[twocolumn,prd,showpacs,10pt,superscriptaddress,nofootinbib]{revtex4}
\usepackage{graphics}
\usepackage{amsmath, amssymb}

\def\MM{{\cal M}}

\def\be{\begin{equation}}
\def\ee{\end{equation}}
\def\bea{\begin{eqnarray}}
\def\eea{\end{eqnarray}}
\def\ba{\begin{array}}
\def\ea{\end{array}}
\newcommand{\lsim}{\,\raise 0.4ex\hbox{$<$}\kern -0.8em\lower 0.62ex\hbox{$\sim$}\,}
\newcommand{\gsim}{\,\raise 0.4ex\hbox{$>$}\kern -0.7em\lower 0.62ex\hbox{$\sim$}\,}

\def\de{\mathrm{DE}}

\def\be{\begin{equation}}
\def\ee{\end{equation}}

\def\deH{\delta{H}}
\def\de F{\delta{F}}

\def\ddH{\ddot{H}}

\def\deH{\delta{H}}
\def\dedH{\delta{\dot{H}}}
\def\deddH{\delta{\ddH}}

\def\delambda{\delta{R}}
\def\desigma{\delta{G}}

\def\H0{H_{0}}

\def\ndmu{\nabla_{\mu}}

\def\nukappa{\nabla^{\kappa}}
\def\ndkappa{\nabla_{\kappa}}

\def\ndnu{\nabla_{\nu}}

\def\nulambda{\nabla^{\lambda}}
\def\ndlambda{\nabla_{\lambda}}

\def\de{\delta}

\def\no{\nonumber}

\def\Ht{\dot{H}}
\def\Htt{\ddot{H}}

\def\Fn{F}

\def\xisn{\xi_{G}}
\def\Fsn{F_{G}}
\def\xiln{\xi_{R}}
\def\Fln{F_{R}}

\def\Fnt{\dot{F}}
\def\xint{\dot{\xi}}
\def\xisnt{\dot{\xi}_{G }}
\def\Fsnt{\dot{F}_{G }}
\def\xilnt{\dot{\xi}_{R }}
\def\Flnt{\dot{F}_{R }}

\def\ka{\frac{k^2}{a^2}}

\def\eff{{\rm eff}}

\begin{document}

\title{Anisotropic stress and stability in modified gravity models}
\date{April 7, 2011}

\author{Ippocratis D. Saltas}
\email{I.Saltas@sussex.ac.uk}
\affiliation{Astronomy Centre, University of Sussex, Falmer, Brighton
BN1 9QH, UK}
\author{Martin Kunz}
\email{Martin.Kunz@unige.ch}
\affiliation{D\'epartement de Physique Th\'eorique, Universit\'e de Gen\`eve, 1211 Geneva 4, Switzerland}

\begin{abstract}
The existence of anisotropic stress of a purely geometrical origin seems to be a characteristic of higher order gravity models, and has been suggested as a probe to test these models observationally, for example in weak lensing experiments. In this paper, we seek to find a class of higher order gravity models of $f(R,G)$ type that would give us a zero anisotropic stress and study the consequences for the viability of the actual model. For the special case of a de Sitter background, we identify a subclass of models with the desired property. We also find a direct link between anisotropic stress and the stability of the model as well as the presence of extra degrees of freedom, which seems to be a general feature of higher order gravity models. Particularly, setting the anisotropic stress equal to zero for a de Sitter background leads to a singularity that makes it impossible to reach the de Sitter evolution.
\end{abstract}

\keywords{cosmology: dark energy}
\pacs{04.50.Kd, 98.80.-k, 95.36.+x}  
\maketitle

\section{Introduction}

Even though it is now well over ten years since supernova data
\cite{sn1,sn2} led to the general acceptance amongst cosmologists
that the expansion of the Universe is accelerating, there are
no natural models that could explain this phenomenon. The nature of the postulated
dark energy which is responsible for the accelerated expansion is still a mystery \cite{Copeland_Review_DE,Peebles&Ratra_DE_Review,FriemanEtal_DE_Review,Perivolaropoulos_DE_Review,Padmanabhan_DE_Review,Durrer&Maartens,sapone_review}.
This fact, as well as expectations from candidates for the
``theory of everything'', like string theory, have led researchers to
investigate models in which General Relativity (GR) itself is modified \cite{Nojiri&Odintsov_Review,Capozziello&Francaviglia_f(R)_Review,f(R)SotiriouFaraoni,f(R)Felice&Tsujikawa,Nojiri&Odintsov_R+f(G),Nojiri&Odintsov_GB_DE,Caroll_etal,Nojiri&Odintsov_GB_DE2,Capozziello_Review}. One
crucial question therefore is whether it is possible to distinguish
between a dark energy ``field'' or fluid on the one hand, and an accelerated
expansion due to a modification of the theory of gravity on the other hand.

Although strictly speaking cosmological probes cannot provide conclusive
proof \cite{ks,kas,husa}, there are certain signatures that can point the
way. One of them is the presence of a significant anisotropic stress:
canonical scalar fields do not create additional anisotropic stress, while
the modified-gravity (MG) models like scalar-tensor theories, brane-world
models like the Dvali-Gabadadze-Porrati (DGP) model \cite{DGP} and $f(R,G)$
type theories generically induce a large effective anisotropic stress.

In this paper we investigate one specific class of models, $f(R,G)$ type
modifications of GR, and ask the question whether it is possible to construct
viable models with a vanishing, or arbitrarily small effective anisotropic stress. Or in other words,
is it possible to mimic ``GR'' with these models, at least up to first order in perturbation
theory and in the sense that
the extra anisotropic stress is small enough? Since $f(R)$ models have many things
in common with scalar-tensor theories, we expect that our discussion is also
relevant for those models, and as we discuss later, also for DGP and other
braneworld models.

We structure the paper as follows: The following section serves to define
our notation and introduces the main equations. In Section \ref{sec:EffectiveAnisStress} we discuss the notion of anisotropic stress in general, and how this plays an important role in modified gravity models and then we investigate the 
possibility of a vanishing anisotropic stress in the particular cases of $f(R)$ and $f(G)$ models, before we look at the more general $f(R,G)$ case. In Section \ref{sec:dSStability} we identify and discuss the link between anisotropic stress and stability in modified gravity models in the context of both homogeneous and inhomogeneous perturbations around de Sitter space. We further derive the relevant stability conditions. We generalize the discussion to arbitrary backgrounds in Section \ref{sec:matter} and give some results for a matter dominated evolution. In Section \ref{sec:ToyModels}, we apply the above to characteristic toy models, and then discuss our conclusions. Some explicit intermediate calculations and formulas can be found in the Appendices.


\section{Action and equations of motion}

We are interested in the class of models described by the action
\be
S = \int_{\MM} d^{4}x \sqrt{-g} \left[ f(R,G) + L_{\rm matter}  \right], \label{eq:action1}
\ee
where $R$ and $G$ are the Ricci and Gauss--Bonnet scalar respectively, $\MM$ denotes the four dimensional spacetime, and $L_{\rm matter}$ is the Lagrangian for any matter fields or fluids present. ``$f$" is an analytic function and its form is constrained by both classical and quantum stability requirements as well as agreement with both large scale and solar system data. (We will revisit this point in Section \ref{sec:dSStability}.) In the following, we will work in natural units where $8 \pi G = c = 1$, unless otherwise stated. 

The Gauss--Bonnet term, which is a topological invariant in four dimensions, is defined as
\be
G \equiv R^2 - 4R^{\mu \nu}R_{\mu \nu} + R^{\mu \nu \kappa \lambda}R_{\mu \nu \kappa \lambda}, \label{GBDefinition}
\ee
and belongs to the so-called Lovelock scalars \cite{Lovelock}, which is the first term in a series of non-linear corrections to the Einstein-Hilbert term.

The class of models described by action (\ref{eq:action1}) are a natural generalisation of the well-studied $f(R)$ models \cite{Nojiri&Odintsov_Review,Capozziello&Francaviglia_f(R)_Review,f(R)SotiriouFaraoni,f(R)Felice&Tsujikawa}. The introduction of curvature scalars other than the Ricci one has been physically motivated by the lower energy limit of string theory \cite{ZwiebachLowEnergyLimit}, although care must be taken to avoid generic instabilities (e.g. \cite{woodard}).
It is well known that actions like (\ref{eq:action1}) can be expressed in an equivalent way as a scalar-tensor action in the so-called Jordan frame, however, in this paper we will work in the original representation. 

Varying action (\ref{eq:action1}) with respect to the metric $g^{\mu \nu}$, and using the Bianchi identities, we get the equations of motion \cite{Carrol_ModGrav,FeliceSuyama}
\begin{align}
&F \left( R_{\mu \nu} - \frac{1}{2} g_{\mu \nu} R \right)  = T_{\mu \nu}^{\rm{(matter)}} + T_{\mu \nu}^{\rm{(eff)}}, \label{Eom}
\end{align}
where $T_{\mu \nu}^{\rm{(eff)}}$ is an effective energy-momentum tensor of geometrical origin (in contrast to the ordinary matter one) and is defined as
\begin{align}
&T_{\mu \nu}^{\rm{(eff)}} \equiv - \Big( \ndmu \ndnu F - g_{\mu \nu}\Box F + 2R\ndmu \ndnu \xi \no \\
& - 2g_{\mu \nu}R \Box \xi - 8R_{(\mu}{}^{\kappa} \ndkappa \nabla_{\nu)} \xi  + R_{\mu \nu}\Box \xi \no \\
& + 4g_{\mu \nu}R^{\kappa \lambda}\ndkappa \ndlambda \xi + 4R_{\mu \kappa  \lambda \nu} \nukappa \nulambda \xi - \frac{1}{2} g_{\mu \nu} V(R,G)\Big), \label{EffectiveEnergyMomentum}
\end{align}
where we used the additional definitions \footnote{Here we follow the notation of \cite{FeliceSuyama}.}
\be F \equiv f_{R} \equiv  \frac{\partial f(R,G)}{\partial R}, \ee
\be \xi \equiv f_{G} \equiv \frac{\partial f(R,G)}{\partial G},\ee
\be V(R,G) = RF + \xi G - f(R,G) . \ee
Taking the limits $\xi \rightarrow 0$ and $F \rightarrow 1$ in (\ref{Eom}), we recover the $f(R)$ and $R + f(G)$ equations of motion respectively.  

We can formally recover a GR-like equation of motion by moving additionally $(F-1) (R_{\mu \nu} - \frac{1}{2} g_{\mu \nu} R)$ to the right hand side of Eq.~(\ref{Eom}) and adding this to $T_{\mu \nu}^{\rm{(eff)}}$ to form a total effective ``dark energy'' energy-momentum tensor, $T_{\mu \nu}^{\rm{(eff, total)}}$. We can then compute effective fluid quantities, for example the equation of state, pressure perturbation and anisotropic stress, that a dark energy needs to have to lead to the same cosmological observations as the original $f(R,G)$ model. But strictly speaking, $T_{\mu \nu}^{\rm{(eff, total)}}$ although covariantly conserved, is not an energy-momentum tensor in the usual sense, since it is a function of the spacetime geometry and its first and second derivatives.

In the following, we will be interested in homogeneous, isotropic and flat cosmologies, described by the flat, four dimensional FRW metric
\be
ds^2 =-dt^2 + a(t)^2 d\mathbf{x}^2,
\ee
with $a(t)$ the scale factor.
In this background, the two key quantities, $R$ and $G$, can be expressed purely as a function of the Hubble parameter $H\equiv H(t)$ and its time derivative,
\bea
R(t) &=& 6 \left( 2 H^2 + \dot{H} \right), \label{R(t)} \\
G(t) &=& 24 H^2 \left( H^2 + \dot{H} \right) \label{G(t)}.
\eea
The t--t component of the $f(R,G)$ equations of motion (\ref{Eom}) gives a modified version of the usual Friedman equation which reads as
\be
H^2 = \frac{1}{F} \left(\frac{1}{3}T^{{\rm (mat)}0}{}_{0} - H\dot{F} - 4H^{3}\dot{\xi} + \frac{1}{6}V \right), \label{f(R,G)FriedmanEquation}
\ee
with dots denoting differentiation with respect to cosmic time $t$. Notice that the above equation is of fourth order with respect to the scale factor, in contrast to the usual Friedman equation.


\section{The effective anisotropic stress in higher order gravity} \label{sec:EffectiveAnisStress}

Let us here introduce the notion of anisotropic stress in gravity. As a starting point, we consider scalar perturbations around a flat FRW background in the conformal Newtonian gauge, where the metric is of the form\footnote{The general form of the perturbed line element is given in Appendix \ref{sec:InhomPertApp}.}
\be
ds^2 = -(1 + 2\psi)dt^2 + a(t)^2\left( 1 -  2\phi \right)d{\bf x}^2, \label{Perturbed_Metric_Newt}
\ee
and the gravitational potentials $\psi \equiv \psi({\bf x}, t)$ and $\phi \equiv \phi({\bf x}, t)$ are closely related to observations: light deflection is sourced by the lensing potential $\phi+\psi$ and non-relativistic particle motion by $\psi$ alone. 

The {\it scalar anisotropic stress} $\Pi$ is then defined as the difference in the potentials 
\be
\phi - \psi \equiv \Pi({\bf x}, t) , \label{Anisotropy_Equation1}
\ee
or the difference of the relevant potentials in some other gauge. Equation (\ref{Anisotropy_Equation1}) is called the {\it anisotropy equation}, and can be found by calculating the $ij$ ($i \neq j$) component of the perturbed equations of motion around the FRW metric,
\bea
\delta G^{i}{}_{j}  - \frac{1}{3}g^{i}_{j} \delta G^{\kappa}{}_{\kappa} &=& \delta T^{({\rm eff, total})i}{}_{j}  - \frac{1}{3} g^{i}_{j} \delta T^{({\rm eff, total}) \kappa}{}_{\kappa} \nonumber \\
&\equiv& \Pi^{({\rm eff})i}{}_{j}  ,
\eea
from which one then extracts the scalar part as usual to get
\be
\phi - \psi = \Pi^{({\rm eff})}. \label{Effective-Aniso-1}
\ee
We emphasize that this is the anisotropic stress one would infer by assuming GR to hold, not only the anisotropic stress from the matter fields. Indeed, here we are precisely interested in the contribution to $\phi-\psi$ due to a modification of gravity. While relativistic particles do induce an anisotropic stress, it is small at late times and we will neglect the contribution of $T_{\mu \nu}^{\rm{(matter)}}$ in equation (\ref{Eom}) to $\phi-\psi$. 
Notice that because of the nature of $T^{({\rm eff})i}{}_{j}$ in modified gravity theories, the r.h.s of above equation will in principle have a spacetime dependence, i.e it will be a function of $\phi, \psi$ as well as their first and second derivatives with respect to time (in Fourier space), in contrast to GR, where the r.h.s is just a function of the matter content. The usefulness of (\ref{Effective-Aniso-1}) is that it has a GR-like l.h.s., allowing to compute predictions for cosmological observations as usual, while all the extra contributions are moved to the r.h.s. and interpreted as a ``modified gravity energy-momentum tensor''.

In particular, for GR (and neglecting any relativistic species) we have $\Pi^{({\rm eff})} = 0$ and therefore $\phi = \psi$ {\it at all times}. Therefore, the inequality of the Newtonian potentials is a ``signature" of departures from GR on large scales \cite{ks}. The ratio $\phi/ \psi$, or variables derived from it, like $\eta(t, k) \equiv \frac{\psi}{\phi} - 1$, can be extracted observationally by combining weak lensing experiments with e.g. galaxy surveys or redshift space distortions, making cosmological observations a powerful test of GR \cite{aks}. Current limits are rather weak, with deviations of order unity from $\eta=0$ still allowed, but future probes will measure the ratio $\psi/\phi$ with an accuracy of a few percent (e.g.~\cite{bean,daniel,song1,song2,lukas}).

In this paper, we raise and investigate the following question: Can we construct a viable modified gravity model with $\phi/\psi = 1$, or in other words, is $\phi\neq\psi$ an unavoidable consequence of modifying gravity to explain the dark energy? We will try to answer this question step by step, by investigating the anisotropy equations of $f(R)$, $R+f(G)$ as well as of the more general $f(R,G)$ gravity models.

The equations for general spaces tend to be complex and in general do not admit simple solutions. For this reason in this paper we will first focus on the case of a de Sitter background. On the one hand, solutions that explain the observed accelerated expansion usually tend towards a de Sitter fixed point, and also the observed background expansion requires no deviation from $p=-\rho$ for the inferred dark energy component. On the other hand, the equations simplify significantly in this limit, which allows us to give explicit solutions that we can then discuss in detail. We comment on the behavior for other backgrounds in section \ref{sec:matter}, but leave a fully general study for future work. We nonetheless expect our conclusions to be quite generic for models that try to explain the dark energy.


\subsection{The anisotropic stress in $f(R)$ models\label{sec:fR}}

Let us begin with the special case of $f(R)$ gravity, described by the action
\be
S = \int_{\MM} d^{4} x  \sqrt{-g} f(R),
\ee
which corresponds to the limit of $\xi \rightarrow 0$ of the general $f(R,G)$ models. It is well known that these models are characterized by an extra, dynamical scalar degree of freedom $F$, which is proportional to the first derivative of $f(R)$, $F \equiv f_{R}(R) \equiv f'(R)$. This can be explicitly seen by taking the trace of the corresponding equations of motion and arriving at a Klein--Gordon type equation for $F$. However, unless the theory is written in the so--called Jordan frame, the latter degree of freedom is still of geometrical origin.

The expression for the anisotropic stress for the general case of $f(R,G)$ gravity is given in equation (\ref{f(R,G)AnisotropyEq}). For $f(R)$ gravity, in the Newtonian gauge, it takes the form
\be
\phi - \psi =  \frac{\delta F}{F} \equiv \Pi^{({\rm eff})}_R, \label{f(R)-Aniso}
\ee
which holds for any spacetime, not just de Sitter. 

Since $\delta F = f_{RR}(R) \delta R$, the stress contribution is proportional to the derivative of the extra scalar degree of freedom with respect to $R$, that is, it depends on the evolution of the scalar $F \equiv f'(R)$. Seeking a form for the function $f(R)$ that would make $\Pi^{\rm (eff)}$ vanish at all times corresponds to solving the equation $f_{RR} = 0$ with a general solution $f(R) = R + \Lambda$, i.e. of all $f(R)$ models it is {\em precisely} GR that satisfies this equation. In other words, the requirement of zero anisotropic stress in $f(R)$ theories is equivalent to suppressing the extra degree of freedom of the theory, leading to the GR limit. (In the Parametrized Post-Friedmannian (PPF) framework of Ref. \cite{husa}, $f_{RR} \rightarrow 0$ corresponds to $B \rightarrow 0$, $B$ being a parameter introduced to quantify the modification from GR).

Although it is not possible to make $\Pi^{\rm{(eff)}}$ exactly zero at all times without reverting back to GR, one can try to make it sufficiently small for a given cosmological period, by an appropriate choice of the model parameters. This corresponds to setting $f_{RR}$ sufficiently small for some particular initial conditions and ensuring that it stays small, by an approriate choice of model. This has been done for example in Ref. \cite{Pog_Pattern}. The price one pays is a rapid oscillatory behavior for both the gravitational potentials and the curvature perturbation. What is more, the amplitude of the latter can grow arbitrarily as $f_{RR} \rightarrow 0$. We will come back to this this later, when we will study the relevant stability conditions 
and will see that this is a general feature of $f(R,G)$ and other modified gravity models: the existence of anisotropic stress is related to the extra scalar degree of freedom of these models, and an attempt to suppress it causes unstable behavior. In the $f(R)$ case, suppression of the extra scalar corresponds to $f_{RR} \rightarrow 0$.
 
As can be seen from equation (\ref{f(R)-Aniso}), another way to force $\Pi^{({\rm eff})}_R = 0$ would be to impose the condition $\delta R = 0$. The crucial difference between $\delta R =0$ and $f_{RR} = 0$, is that the latter is a background requirement, i.e a requirement on the particular form of the $f(R)$ action. On the other hand, the condition $\delta R = 0$ imposes a dynamical condition on the potentials $\phi, \psi$ and their first and second time derivatives.  If we also take into account that in that case the l.h.s implies $\phi = \psi$, we find the equation
\be
\ddot{\phi} + 5H\dot{\phi} + 3H^2 \left( \frac{H^2}{\dot{H}} + \frac{k^2}{6H^2a^2} + 2 \right)\phi = 0,
\ee
which not only in general is unstable, but also fixes the perturbation evolution needed to keep $\delta R=0$, which is in general incompatible with the desired evolution of the universe, e.g. structure formation. In other words, the requirement $\delta R=0$ {\em imposes} an evolution that in general we do not want. For this reason, what we seek in this paper is a condition of the first kind, i.e. a condition on model space rather than on the evolution of the perturbations.

\subsection{The anisotropic stress in $f(G)$ models\label{sec:fG}}

Since $f(R)$ models do not allow for a vanishing anisotropic stress, we will instead look at the other limiting case of $f(R,G)$ models, namely those described by the action
\be
S = \int_{\MM} d^{4} x  \sqrt{-g}  \left[ R+ f(G) \right].
\ee
These models posses an instability in the presence of a matter fluid, irrespective of the form of the function $f(G)$ \cite{Fel_GBInstab}, which rules them out as realistic scenarios, but here we just want to see whether it is possible to construct $f(G)$ models that contribute no additional effective anisotropic stress.

The first term in the action does not contribute any extra anisotropic stress. In an FRW background, these models posses an extra scalar degree of freedom, proportional to $\xi \equiv f_{G}(G) = f'(G)$. The anisotropy equation in a general spacetime in this case reads as
\be
\phi - \psi  \equiv  \Pi^{({\rm eff})}_{G}  =  4H \dot{\xi}\psi  - 4\ddot{\xi}\phi + 4 \left( H^2 + \dot{H} \right)\delta \xi , 
\ee
with $\delta \xi = f_{GG}\delta G$. For a de Sitter background, the equation simplifies to $\phi - \psi  = 4 H_0^2 f_{GG} \delta G$. One possibility to have no anisotropic stress is to set $f_{GG}=0$ at all times, leading to the model $f(G) = G + \Lambda$. In four dimensions $G$ is a topological invariant \cite{Lovelock}, i.e. it is a total derivative and so it has no contribution to the equations of motion, and we are left only with $R + \Lambda$ for the relevant gravitational Lagrangian, which is equivalent to GR. Alternatively we require $\delta G = 0$ which suffers from the same problems as $\delta R = 0$ and does not allow in general for a sensible evolution of the perturbations.

For a general background, the similarity to the case of $f(R)$ is spoiled by the first two terms in the anisotropy equation. In general, the condition on the evolution of $\phi$ and $\psi$ imposed by those terms will again be difficult to enforce as a function of time. On the other hand, if the background quantities vary only slowly,  $ \dot{\xi}, \ddot{\xi} \approx 0$, then the anisotropy equation can be simplified as \footnote{We obtain effectively the same condition on scales that are well inside the horizon, $k\gg a H$, as $\delta\xi$ is in general boosted by factors of $(k/(aH))^2$ relative to $\phi$ and $\psi$.}
\be
\phi - \psi  \equiv  \Pi^{({\rm eff})}_{G}  = -4\left( 1 + 3w_{\rm eff} \right)\delta \xi,
\ee 
where we used the relation 
\be 
\frac{\dot{H}}{H^2} \approx - \frac{3}{2}(1 + w_{\rm eff}), \label{EffectiveExpansion}
\ee
with $w_{\rm eff} \equiv p/ \rho$ being the effective equation of state parameter for the background evolution. Now, the situation is again similar to the one encountered for $f(R)$: one has either to require either $f_{GG} = 0$, $\delta G =0$, or $w_{\rm eff} = -1/3$. As discussed above, the first condition leads to GR (in which case automatically $\dot{\xi}=\ddot{\xi}=0$ at all times), while the second does not allow for an acceptable evolution of the perturbations. The third condition, which corresponds to the evolution of a universe dominated by curvature, is also not very relevant given current observational results in cosmology.


\subsection{The anisotropic stress in $f(R,G)$ models \label{sec:fRG}}

We saw in the previous sections that the vanishing of the anisotropic stress in $f(R)$ and $f(G)$ models corresponds to either trivial or unphysical situations. We now turn to study
the more general case of $f(R,G)$ models. Here, the function $f(R,G)$ has two contributions, coming from the $R$- and $G$- part respectively, and the anisotropy equation reads us
\begin{align}
\phi  -  \psi & = \frac{1}{F}  \left[  \delta F +  4H \dot{\xi}\psi  - 4\ddot{\xi}\phi + 4\left( H^2 + \dot{H} \right)\delta \xi\right].   \label{AnisoEqf(R,G)General}
\end{align}  
Unlike the $f(R)$ case, where we simply had to demand that $f_{RR}(R)=0$, the nature of the anisotropy equation here does again not allow us to write down an explicit condition for the function $f(R,G)$ that would give a zero anisotropic stress contribution in a general spacetime: as in $f(G)$ models, we find extra factors of $\phi$, $\psi$ and their time derivatives. The only case for which we can find a simple condition is for the de Sitter spacetime, and therefore we shall restrict ourselves in this case for the time being. Furthermore, for models that try to explain the dark energy, it is at late times that we expect modifications of gravity to become important, and that deviations from GR should appear in observations. For such a late-time accelerating epoch, a de Sitter spacetime is expected to provide a reasonable approximation.

The anisotropy equation in de Sitter space reads as
\begin{align}
\phi  -  \psi & = \frac{1}{F}  \left[  \delta F + 4H_{0}^2 \delta \xi  \right]  \no \\
&\equiv \Pi^{({\rm eff})}_{G} +  \Pi^{({\rm eff})}_{R} \equiv \Pi^{({\rm eff})}_{\rm tot}, \label{f(R,G)AnisotropyEq}
\end{align}
where, as before, we have defined the contribution coming from the $R$- and $G$- part of the action respectively as
\be
\Pi^{({\rm eff})}_{R} \equiv  \frac{\delta F}{F} \; \; \; {\rm and} \; \; \; \Pi^{({\rm eff})}_{G}  \equiv  4H_{0}^2  \frac{\delta \xi}{F}. \label{Eff_Stress_f(R)_f(G)}
\ee
Notice that this case is just the sum of the corresponding limiting cases of $f(R)$ and $R+f(G)$ gravity respectively,
although now either term depends on both $R$ and $G$.

We now ask the same question as before: Is it possible in this case to find a class of $f(R,G)$ models that give a zero anisotropic stress $\Pi^{({\rm eff})}_{tot} = 0$, having at the same time a sensible evolution of the perturbations? By inspection of (\ref{Eff_Stress_f(R)_f(G)}) one can see that in order for the total scalar anisotropic stress to be zero, we require that at all times 
\begin{align}
\Pi^{({\rm eff})}_{R} = - \Pi^{({\rm eff})}_{G}. \label{StressEquality}
\end{align}
In other words, we require that the particular anisotropic stress contributions have equal magnitude and opposite sign at all times, or at least for the cosmological era of interest. 

We can rewrite condition (\ref{StressEquality}) using the relations
\begin{align}
&\delta F= F_{R}(R, G) \delta R+ F_{G}(R, G) \delta G, \label{deltaF} \\
&\delta \xi= \xi_{R}(R, G) \delta R + \xi_{G}(R, G) \delta G \label{deltaxi}.
\end{align}
In de Sitter space we have additionally 
\be
 G = 4 H^{2}_{0}  R,
\ee
which implies that $\delta G = 4 H^{2}_{0} \delta R$. Using the last relation together with (\ref{deltaF}) and (\ref{deltaxi}) (and so limiting ourselves to de Sitter backgrounds) condition (\ref{StressEquality}) becomes
\begin{align}
 \left( f_{R R}+4H_{0}^2 f_{R G} + 4H_{0}^2 f_{G R } +16H_{0}^4 f_{G G } \right) \delta R = 0. \label{Zero_Stress_Diff_Eq_0}
\end{align}
If $f(R, G)$ is an analytic function we have $f_{R G } = f_{G R }$, and requiring that the above equation
is valid for any variation $\delta R$ (see the discussion on $f(R)$ in Section \ref{sec:fR}) we arrive at
\begin{equation}
f_{R R} + 8 H_{0}^2 f_{R G} + 16 H_{0}^4 f_{G G} =0. \label{Zero_Stress_Diff_Eq}
\end{equation}
The above equation is a second order partial differential equation with constant coefficients, for the class of functions $f \equiv f(R, G)$ that give a vanishing anisotropic stress in de Sitter space. Its general solution is
\begin{equation}
f(R,G) = f_{1} \left( \Omega  \right)+  R f_{2}\left( \Omega \right), \label{Zero_Anis_Stress_Models}
\end{equation}
with $\Omega \equiv  R - G/(4H_{0}^2)$, and $f_{1}$, $f_{2}$ arbitrary but analytic functions of $\Omega$. Models with vanishing anisotropic stress in de Sitter space need to be of this form locally near the de Sitter point.

We specify the function $f(R,G)$ in the action, which is agnostic of quantities like $H_0$. For this reason it is preferable to consider a more general class of models with
\be
\Omega \equiv \left( R - \frac{G}{M^2} \right), \label{Omega_Model_General}
\ee
with $M$ a parameter with mass dimensions, so that $\Pi^{({\rm eff})} \rightarrow 0$ corresponds to the special case of a model with a de Sitter expansion rate of $H_0=M /2$. As we will also discuss later on, the mass parameter $M$ controls which of the two contributions in $f(R,G)$ dominates.

Assuming that the de Sitter point exists and is stable, we see that it is in principle possible to find a non-trivial class of $f(R,G)$ models that give exactly zero anisotropic stress in de Sitter space at all times, by selecting a model in the class (\ref{Zero_Anis_Stress_Models}). However, as we will see by studying the stability of de Sitter space below, the case $M \rightarrow 2 H_{0}$ corresponds to a singularity for the actual model, and therefore the model cannot be viable. Furthermore, we will see that the anisotropic stress cannot become arbitrarily small, since this will cause unstable behavior for the curvature perturbations.



\section{Anisotropic stress and stability for a de Sitter background} \label{sec:dSStability}
There are different stability criteria that a gravitational theory aiming to describe the late time acceleration should satisfy, each leading to a different condition for the form of the function $f(R,G)$.
At the background level, a viable model should give rise to sufficiently long radiation and matter eras, as well as a transition to a stable de Sitter era \cite{Amend_f(R)Conditions,Odints_f(R)Conditions,Zhou_f(G)Conditions}. Furthermore, avoidance of singularities and of rapid collapse of perturbations (positivity of the sound speed) as well as agreement with local gravity constraints should be ensured \cite{Dav_Solarf(G),Amend_SolarGB,Fel_Solarf(G)}. Of great importance is also the absence of ghost like degrees of freedom \cite{Fel_Ghosts,Chi_Ghost,Nun_Ghost}. For the class of $f(R,G)$ models in a de Sitter background the latter requirement translates into $f_{R}(R,G)> 0$. 

Modified gravity models of the type $f(R)$ or $R+f(G)$ suffer from a curvature singularity at very early times of the cosmological evolution \cite{Staro_Sing,Tsuj_f(R)Pert,Fro_Sing,Fel_f(G)Construction,Sot_CurvInstab}\footnote{Other types of singularities in the context of modified gravity have been earlier observed in Ref. \cite{Odin_singularity1}. For a discussion on various type of singularities in the same context see for example Ref. \cite{Odin_singularity2} and references therein.}. The latter singular behavior can lead to oscillations of the scalar degree of freedom with infinite amplitude and frequency. As explained in Ref. \cite{Fro_Sing}, the singularity lies at a finite field value and energy level and therefore is easily accessible. We will see in the following that this singularity is a feature of $f(R,G)$ models as well.
 
In this paper we are interested in the classical stability, and particularly its connection to the effective anisotropic stress. 
As we will show and discuss below, the attempt of turning off or making sufficiently small the effective anisotropic stress for a de Sitter background leads to serious stability problems that question the actual viability of models with vanishing $\Pi^{({\rm eff})}$.
 
 \subsection{Existence of a de Sitter point}
Since we will specifically study the behaviour near the de Sitter point, it is necessary that this solution exists for the models of interest. De Sitter space is a vacuum, maximally symmetric space described by the conditions 
\be
H = H_{0} = {\rm constant} > 0, \; \; \; \dot{R} = \dot{G} = \dot{F} = \dot{\xi} = 0. \label{deSItterCond}
\ee
Furthermore, in maximally symmetric spaces any curvature invariant can be expressed as a function of the Ricci scalar, and particularly for the Gauss--Bonnet term we get
\be
G = \frac{R^2}{6}  . \label{deSitterG}
\ee
We can derive the condition for the existence of the de Sitter point by taking the trace of the equations of motion (\ref{Eom}) and using relations (\ref{deSItterCond}), (\ref{deSitterG}), to arrive at
\be
F(R)R + 2G(R)\xi(R) - 2f(R) =0, \label{deSitterPointCondition}
\ee
where everything is assumed to be expressed in terms of the Ricci scalar and evaluated on de Sitter space. The cases $\xi = 0$ and $F = 1$, give the relevant conditions for $f(R)$ and $R+f(G)$ gravity respectively. Solving the algebraic equation given above, we get the de Sitter point solution,  which in general is not unique. Minkowski space corresponds to the special case of $R_{0} = H_{0} = 0$. 

For the models of the type (\ref{Zero_Anis_Stress_Models}), we find with the help of Eq. (\ref{deSitterPointCondition}) that the de Sitter point is given by solutions of the equation
\be
f_1(u) + u  f_2(u) = 0 .
\ee
and $R=2 u$. The next step in our analysis will be the study of the stability of de Sitter space at both homogeneous and inhomogeneous level.

\subsection{Homogeneous perturbations}\label{sec:dSStabilityB}
Now we turn to study the stability of the de Sitter solution, first with respect to homogeneous (background) perturbations. As we will see, there is a strong link between effective anisotropic stress and stability in modified gravity models. 

Let us consider the $t-t$ component of the Friedman equation (\ref{Eom}) and perturb it linearly around the de Sitter solution $H = H_{0}$ 
\be
H(t) = H_{0} + \delta H(t). \label{dS_Perturbation}
\ee
Under perturbation (\ref{dS_Perturbation}) the perturbed function $f(R,G)$ reads as
\be
f = f_{0}+F_{0}\de R + \xi_{0} \de G, \label{f(R,G)_Pert}  
\ee
and similar expressions hold for the other quantities of interest. The explicit formulas and calculations for any space can be found in Appendix \ref{sec:HomogPert}.

Now, evaluating relations (\ref{C1general})--(\ref{C3general}) and using conditions (\ref{deSItterCond}), 
we can write the linearized perturbed modified Friedman equation (\ref{f(R,G)FriedmanEquation}) in the form
\be
C_{1}\deddH + C_{2} \dedH + C_{3} \deH  = 0, \label{f(R,G)dSStabilityEquation}
\ee
with the constants $C_1, C_2$ and $C_3$ defined in Appendix \ref{sec:HomogPert}. There is no constant term since we know that de Sitter, $\deH\equiv 0$, is a solution. This equation then admits an exponential solution of the form
\be
\delta H = A_{+} e^{a^{+}t} + A_{-} e^{a^{-}t}, 
\ee
with 
\be
a^{\pm} \equiv -\frac{3}{2} H_{0} \pm \sqrt{\frac{9}{4}H_{0}^2 - \left( \frac{F}{3\omega} -4 H_{0}^2 \right)}, \label{BackgroundDeSitterEigenvalues}
\ee
and
\be
\omega \equiv F_{R} + 8H_{0}^2 \left(  F_{G} + 2H_{0}^2 \xi_{G}  \right), \label{omega}
\ee
where we dropped the subscript ``0" from $F_{R}$ e.t.c for simplicity.

From solution (\ref{BackgroundDeSitterEigenvalues}), we can read off the condition for de Sitter space stability with respect to homogeneous perturbations:
\be
\frac{F}{3 \left[ F_{R} + 4H_{0}^2 \left( 2 F_{G} + 4H_{0}^2 \xi_{G} \right) \right]} - 4H_{0}^2 \geq 0. \label{dSStabilityCondHomogf(R,G)}
\ee 
The latter condition ensures that the de Sitter point is an attractor for the particular $f(R,G)$ model under study, which is important for the viability of a cosmological model of gravity. The limit $\xi \rightarrow 0$ in (\ref{dSStabilityCondHomogf(R,G)}) gives the corresponding condition for $f(R)$ gravity, that has been derived before in Ref. \cite{Faraoni_Stabilityf(R)},
\be
\frac{F}{3F_{R}} - 4H_{0}^2 \geq 0, \label{dSStabilityCondHomogf(R)}
\ee
while when $F \rightarrow 1$ we get a similar condition for the $R+f(G)$ models also derived in Ref. \cite{Fel_f(G)Construction}
\be
\frac{1}{48H_{0}^4 f_{GG}} - 4H_{0}^2 \geq 0. \label{dSStabilityCondHomogf(G)}
\ee

The stability condition (\ref{dSStabilityCondHomogf(R,G)}) is general, but now we can check what it tells us for the class of models that give a zero anisotropic stress, described by equation (\ref{Omega_Model_General}) as $M^2 \rightarrow 4H_{0}^2$. We can see that in this case necessarily $\omega \rightarrow 0$, and so the eigenvalues (\ref{BackgroundDeSitterEigenvalues}) tend to infinity\footnote{Since the no-ghost condition requires that $F > 0$, the question whether the background solution moves towards or away from de Sitter depends on whether $\omega \rightarrow 0^{+}$ or $\omega \rightarrow 0^{-}$.}. In particular, when $\omega$ is exactly zero, which corresponds to the case of a vanishing anisotropic stress, it is not possible to reach the de Sitter state without triggering a singularity in the model: the quantity $C_1=18 H \omega$ in Eq. (\ref{f(R,G)dSStabilityEquation}) goes to zero as we approach de Sitter, together with $C_2 \rightarrow 0$ (see appendix \ref{sec:HomogPert}). In general $\deddH \sim (C_3/C_1) \delta H \rightarrow \infty$ which requires $\deddH$ to diverge in order to satisfy the evolution equation, except possibly for a lower dimensional and thus infinitely fine-tuned set of trajectories in specific models. We will show and discuss this explicitly in section \ref{sec:ToyModels} considering examples for particular $f(R,G)$ models.

Additionally, if the effective anisotropic stress is not exactly zero, but very small, then the rapid background oscillations lead to potentially large time derivatives of $\delta H$, which makes the linear analysis unreliable, i.e the evolution becomes non linear. We give an explicit example in section \ref{sec:TMA}.

Similarly, in the $f(R)$ and $R+f(G)$ cases, where the zero anisotropic stress condition was that $f_{RR}(R) = 0$ and $f_{GG}(G)=0$ respectively, conditions (\ref{dSStabilityCondHomogf(R)}) and (\ref{dSStabilityCondHomogf(G)}), give the obvious result that one gets infinities when trying to suppress the extra degree of freedom. The difference with the more general $f(R,G)$ models is that the singularity appears for a finite value of the mass parameter $M$ of the model, while in $f(R)$ and $R+f(G)$ the same happens for rather trivial cases. We conclude therefore that a $f(R,G)$ type model that has no anisotropic stress in a de Sitter background cannot dynamically reach this background solution.

\subsection{Inhomogeneous perturbations}
In this subsection we will study the behavior of inhomogeneous perturbations in de Sitter space and we will first show that the stability condition coincides with the stability condition derived in Section \ref{sec:dSStabilityB} on homogeneous perturbations. This was found to be true for $f(R)$ models in Ref. \cite{Faraoni_Stabilityf(R)} and in general is not true for scalar--tensor models. We will then make the relation between  anisotropic stress and stability clear by studying the evolution of the perturbations.
The full set of perturbation equations together with some useful relations can be found in the Appendix \ref{sec:InhomPertApp}.

We follow \cite{FeliceSuyama} and choose the gauge invariant expression
\be
\Phi \equiv  \frac{1}{2F}  \left[  \delta F + 4H_{0}^2 \delta \xi  \right].
\ee
for the gravitational potential, as it reduces to $\phi$ in the Newtonian gauge and remains well-defined for a de Sitter background.
For that background, we find that the potential is just given by
\be
\Phi = \frac{ \left( f_{RR} + 8H_{0}^2f_{RG} + 16H_{0}^4 f_{GG} \right) \delta R}{2F}, \label{Master_Equation_1}
\ee
where we used the fact that $f_{RG} = f_{GR}$ and that in de Sitter space we have $\delta G = 4H^2 \delta R$. 
From condition (\ref{Zero_Stress_Diff_Eq_0}) we see that in de Sitter space and for models that have no anisotropic stress, $\Phi$ is necessarily zero. However, let us assume that we are not exactly in this limit. 
Then by substituting the expression of $\delta R$ in terms of the gauge invariant $\Phi$, relation (\ref{GI_deltaR}), we arrive at the evolution equation,
\be
\ddot{\Phi} + 3H_{0} \dot{\Phi} + \left( \frac{k^2}{a^2} + m_\eff^2 \right)\Phi = 0, \label{MasterEquation}
\ee
with 
\be
m_{\eff}^2 \equiv \frac{F}{3 \omega} - 4H_{0}^2, \label{EffectiveMass}
\ee
with $\omega$ defined in (\ref{omega}), and $a(t) \propto \exp(H_{0}t)$. $m_{\eff}^{2}$ is the effective mass of the Klein--Gordon type equation for the scalar perturbation in de Sitter space, and has a purely geometrical origin. Equation (\ref{MasterEquation}) reduces to that of $f(R)$ and $R+f(G)$ for the limits of $\xi \rightarrow 0$ and $F \rightarrow 1 $ respectively.

As $k \rightarrow 0$, the requirement for superhorizon stability dictates that the effective mass is positive,
\be
m_{\eff}^{2} > 0 \; , \label{dSStabilityCondInhomog}
\ee
which leads to the same stability condition as derived before with the homogeneous analysis, equation (\ref{dSStabilityCondHomogf(R,G)}). Therefore,  the two stability criteria, with respect to homogeneous and inhomogeneous perturbations respectively, lead to the same conditions, as it is the case for $f(R)$ gravity as well \cite{Faraoni_Stabilityf(R)}. 

Turning back to the effective anisotropic stress, we can see that considering again the class of models found in (\ref{Zero_Anis_Stress_Models}) and requiring $M \rightarrow 2  H_{0}$ ($\Pi^{{\rm (eff)}}_{\rm tot} \rightarrow 0$), will make the denominator of (\ref{EffectiveMass}) go to zero so that
\be
\lim_{M \rightarrow 2 H_{0}} m_{\eff}^{2} \equiv \lim_{\omega \rightarrow 0} \left( \frac{F}{3\omega} - 4H^2_{0} \right) = \pm \infty, \label{EffectiveMassLimit}
\ee
depending on the sign of $\omega$ as it approaches zero. In the case of positive infinity the stability condition is not violated, while the minus infinity will obviously violate the stability condition, as it would make the square of the effective mass negative (tachyonic).

The effective mass going to infinity means that the scalar degree of freedom becomes frozen and so it is effectively suppressed. This is also the case in the special cases of $f(R)$ and $R+f(G)$ gravity, as can be seen by inspection of equations (\ref{dSStabilityCondHomogf(R)}) and (\ref{dSStabilityCondHomogf(G)}) for $f_{RR} \rightarrow 0$ and $f_{GG} \rightarrow 0$ respectively. However, here the singularity appears in a non trivial way, i.e for a critical value of the mass parameter $M$ where the two different contributions, i.e the $R$- and the $G$- contribution in (\ref{StressEquality}) balance each other. By consequence, in $f(R,G)$ type models, the anisotropic stress is related to the extra scalar degree of freedom of the theory. As the same happens in scalar-tensor models (e.g. equation (43) of \cite{aks}), and also in DGP where the absence of anisotropic stress requires the crossover scale to diverge, $r_c \rightarrow\infty$, which effectively restores GR, we conjecture that this is a quite general feature of modified gravity models. In addition, in the $f(R,G)$ case, turning the anisotropic stress off (or trying to make it sufficiently small) has a direct impact on the stability and time evolution of the model.

To see what happens when the mass diverges, it is possible to study the solution of the evolution equation (\ref{Master_Equation_1}) using a WKB approximation. We discuss the procedure in more detail in appendix \ref{app:wkb}, where we show that the solution in this regime, and for a sufficiently large effective mass $m_{\rm eff}$, is approximately given by
\be
\Phi(t) \approx \sum_{\pm} C_{\pm} \exp \left[ - \frac{3}{2}H_{0}t \; \pm \;  i m_{\rm eff} t \right], \label{WKBdeSitterSolutionLimit}
\ee
with $C_{\pm}$ constants and $H_{0} > 0$.
From the above solution it can be seen that the frequency of the oscillations is proportional to $m_{\rm eff} $. Suppressing the anisotropic stress leads to a very large effective mass and thus to a very rapid oscillation of $\Phi$. Although we have shown this here only for the de Sitter limit, we expect that the result is more general, and similar oscillations have been seen for example in Ref. \cite{Pog_Pattern} during matter domination for numerically reconstructed $f(R)$ models which mimic GR at early times. 

From relation (\ref{EffectiveMass}) it can be seen that a sufficiently large effective mass corresponds to a sufficiently small anisotropic stress. However, the curvature perturbation $\delta R$ (or $\delta G$) has an amplitude that is $\propto m_{\rm eff}^2$,
\be
 \delta R(t)  =  6 \left( m^{2}_{\rm eff} + 4H_{0}^2 \right) \Phi(t)
\ee
and as $m^{2}_{\rm eff} \gg 1$ one can get very large curvature perturbations. The latter behavior, occurring while we try to suppress the effective anisotropic stress (unless the initial conditions are tuned appropriately), is very similar to the one caused by the singularity found in Starobinsky's ``disappearing cosmological constant" model, Refs \cite{Staro_Sing} and \cite{Fro_Sing}. In that case, the singularity appeared in the high curvature limit of the particular model, while in our case it appears in the model space of different $f(R,G)$ models respectively. 
The latter oscillatory behavior endangers the stability of the actual model as has been pointed out in Refs \cite{Staro_Sing} and \cite{Fro_Sing}, and for an explicit discussion on the subject the reader is referred to Refs. \cite{Staro_Sing, Fro_Sing}. 

Another interesting aspect of the models of the type $f(\Omega)$ concerns the sound speed. The propagation speed in de Sitter space equals the speed of light ($c^2_{s} = 1$). However, using the formula derived in \cite{FeliceSuyama} we find that the sound speed in a general background is given by
\begin{align}
c^{2}_{s} &= 1 + \frac{8 \dot{H}}{4H^2 - M^2} \equiv 1 + \left( \frac{2}{1 - \gamma} \right)  \frac{\dot{H}}{H^2} \label{SoundSpeed1}
\end{align}
where $\gamma \equiv  \frac{M^2}{4H^2}$ is a dimensionless parameter (constant in a de Sitter background).  $\gamma \gg 1$ implies that the Ricci scalar part of the $f(R,G)$ contribution to the anisotropic stress dominates, while for $\gamma \ll 1$ the Gauss--Bonnet part is larger. $\gamma = 1$ corresponds to the case where the two contributions in $f(R,G)$ models become equal and cancel.

We can calculate $\dot{H}$ from the equations of motion, and for this particular class of models we get
\be
\dot{H} =  \frac{ (1 - \gamma)(H \dot{\xi} - \ddot{\xi}) }{ 8H^2(F + 4H\dot{\xi}) },
\ee
which can then be substituted in (\ref{SoundSpeed1}). However, considering an expansion characterized by an effective $w_{\rm eff}$, equation (\ref{EffectiveExpansion}), the sound speed takes the form
\be
c^{2}_{s} \approx 1 - \frac{ 3 (1+ w_{\rm eff}) }{1 - \gamma}.
\ee
Assuming a background with $w_{\rm eff} \neq -1$, we immediately see that as $\gamma \rightarrow 1$, $c^{2}_{s} \rightarrow \infty$. The sound speed becomes negative for $\gamma < 1$ (Gauss--Bonnet part dominates) and positive for $\gamma > 1$ (Ricci scalar part dominates) respectively. The value $\gamma = 1$, which corresponds to the effective anisotropic stress becoming zero is the critical value where the sound speed diverges and changes sign. In other words, if one wishes to enforce $c_s \leq 1$ then one has to ensure that the model lies sufficiently far from the regime where the two contributions balance (see e.g. \cite{bcd1, bcd2} for a discussion on possible issues of superluminal propagation of perturbations).


\section{General and matter-dominated background\label{sec:matter}}

In this section we extend the analysis to a general background evolution, and then consider specifically the important case of matter domination.  In general we have to consider Eq.~(\ref{AnisoEqf(R,G)General}). In this equation, $\delta F$ and $\delta \xi$ are functions of $\delta R$ and $\delta G$ through Eqs.~(\ref{deltaF}) and (\ref{deltaxi}). These in turn can be expressed in terms of the metric perturbations, $\phi$ and $\psi$, see e.g. \cite{frg_evolution}. In the small-scale limit, $k\gg a H$, we find that $\phi=\psi$ implies
\be
f_{RR} + 16 (H^2+\dot{H}) (H^2 + 2 \dot{H}) f_{GG} +4 (2 H^2+3 \dot{H}) f_{RG} = 0 .  \label{eq:noanisoH}
\ee
In order to re-transform this condition into one involving only $R$ and $G$, we can eliminate $H$ and $\dot{H}$ with the help of equations (\ref{R(t)}) and (\ref{G(t)}),
\bea
H^2 &=& \frac{1}{12} \left(R + \sqrt{R^2 - 6 G}\right) , \\
\dot{H} &=& - \frac{1}{6} \sqrt{R^2 - 6 G} .
\eea
Using this prescription we find for the general no-anisotropic-stress condition
\bea
0 &=& f_{RR} + \frac{2}{9} \left(-9 G + 2 R \left(R-\sqrt{R^2-6 G} \right) \right) f_{GG} \nonumber \\
  && + \frac{2}{3} \left( R - 2 \sqrt{R^2 - 6 G} \right) f_{RG} .
\eea

While it is difficult to find general solutions, we can instead study the case for a background evolving with a given $w_{\rm eff}$, as defined in (\ref{EffectiveExpansion}). We notice that in this case equation (\ref{eq:noanisoH}) can be written as
\bea
0 &=&f_{RR}  - 2 H^2 (5+ 9w_{\rm eff}) f_{RG}  \nonumber \\ &&+ 8 H^4 (2+ 9 w_{\rm eff} (1+w_{\rm eff})) f_{GG} .
\eea
For $w_{\rm eff}=-1$ (de Sitter expansion) we recover Eq.~(\ref{Zero_Stress_Diff_Eq}), while for $w_{\rm eff}=0$ (matter dominated expansion) we find
\be
f_{R R} - 10 H^2 f_{R G} + 16 H^4 f_{G G} =0. \label{Zero_Aniso_Cond_Matter}
\ee
The Hubble parameter in the latter equation can be eliminated in favor of R and G using equations (\ref{R(t)}) and (\ref{G(t)}) evaluated for a matter background,
\be
R = 3H^2, \; \; \; G=-12H^4, \; \; \; G = -\frac{4}{3}R^2. \label{Matter_R_G}
\ee
We now try to construct an explicit example for a model that has no anisotropic stress during matter domination. For this purpose, we make an ansatz
\be
f(R,G) = R +  G^n \beta(R)
\ee
Here we take $\beta$ as an a-priori general function of $R$. Inserting this model into Eq.~(\ref{Zero_Aniso_Cond_Matter}) and using (\ref{Matter_R_G}) we can re-express the condition in terms of $R$ only. We find that $\beta$ needs to satisfy the following differential equation:
\be
2 n (n-1) \beta +5 n R \beta' +2 R^2 \beta'' = 0 .
\ee
This equation has clearly a power-law solution, $\beta(R) = c R^m$, with 
\be
m_{1,2} = \frac{1}{4} \left(2 - 5n \pm \sqrt{4 + n (9n-4)} \right) ,  \label{eq:exp_noaniso}
\ee
and the general solution is of the form 
\be
f(R,G) = R + c_1 G^n R^{m_1} + c_2 G^n R^{m_2} \label{eq:mattertoy}
\ee
where $m_i=m_i(n)$ is given by the equations for $m_1$ and $m_2$ above.

A successful model with zero anisotropic stress should at the same time satisfy the Friedmann equation as well. During matter domination we can write the latter as 
\bea
0 &=& R^2 f_{RR}  - 2G^2 f_{GG}  + RG f_{RG} - \frac{1}{6} R f_{R} + \frac{1}{6} G f_{G} \nonumber \\
&& - \frac{1}{6}f + \frac{3}{4} \rho_0 R \label{Friedman_Matter}.
\eea
Here we chose $R$ and $G$ so as to correspond to the partial derivatives, since the choice is not unique. The final term is due to $\rho_m(t) \propto t^{-2} \propto R$. Inserting a model of the form (\ref{eq:mattertoy}) but for a general exponent $m$, we find the condition
\be
-6 m^2+m (7-6 n)+(n-1) (12 n-1)=0 . \label{eq:exp_matdom}
\ee
A model of this form that satisfies simultaneously (\ref{eq:exp_noaniso}) and (\ref{eq:exp_matdom}) allows for a matter dominated evolution and contributes no anisotropic stress during that period. This is the case for
\be
n = \frac{1}{90} \left( 11 \pm \sqrt{41} \right) \; , \quad m = \frac{1}{180} \left( 61 \pm 11 \sqrt{41} \right) \label{eq:mattertoyExponents}
\ee
where one needs to use either both positive or both negative signs. An additional solution is given by $m=0$ and $n=1$, which is just GR.

Therefore, there is at least one model in the context of $f(R,G)$ gravity that is able to give a zero effective anisotropic stress, in the subhorizon limit of a matter background. 

Let us now turn attention to homogeneous perturbations around the matter point, keeping the function $f(R,G)$ in its general form for the start. In Appendix \ref{sec:HomogPert} we calculate the evolution of homogeneous perturbations for a general expansion $a(t) \propto t^{p}$. For the matter case we get for $p = 2/3$
\begin{align}
\delta \ddot{H}  + \left( \frac{\dot{\omega}}{\omega} + \frac{9H}{2}  \right) \delta \dot{H} + m_{\rm \eff}^2 \delta H= \frac{\delta \rho_{\rm m}}{18H \omega}, \label{HomogPertMatter}
\end{align}
with the effective mass defined as
\be
m_{\rm \eff}^2 \equiv \frac{F}{3 \omega} \equiv \frac{F}{3\left[ F_{R} + 4H^2 \left( 2 F_{G} + 4H^2 \xi_{G} \right) \right]}.
\ee

Equation (\ref{HomogPertMatter}) can be solved approximately at the WKB regime using an iterative approach \cite{Staro_Sing, Tsuj_f(R)Pert},
\be
\delta H =  \delta H_{(\rm osc)}  + \delta H_{(\rm ind)}. \label{SoldHomogPertMatterGeneral}
\ee
$\delta H_{(\rm osc)} $ is the solution describing oscillations of the scalar degree of freedom, obtained setting $\delta \rho_{\rm m} = 0$. $\delta H_{(\rm ind)}$ denotes the matter induced part, which is obtained by turning off all the derivatives on the l.h.s of equation (\ref{HomogPertMatter}). We assume that $| \delta H_{(\rm osc)} | \ll \delta H_{(\rm ind)}$, so that the deviations from GR are sufficiently small.

Stability in this case requires, apart from the no--ghost condition $F>0$, that the effective mass is positive, 
\be
m_{\rm \eff}^2 > 0. \label{StabCondMatter}
\ee

Let us turn attention to the oscillatory part of the solution (\ref{SoldHomogPertMatterGeneral}). It can be obtained using the WKB approximation, by assuming the solution is a slowly varying quantity in time,
\be
\delta H_{\rm (osc)} \approx Ae^{i \theta(t)},
\ee
with $\ddot{\theta} \ll 1$. Plugging above ansatz into (\ref{HomogPertMatter}), and after some algebra, we find that
\begin{align}
\delta H_{\rm (osc)} \approx & \sum_{\pm} A_{\pm} \exp \left[ -\frac{1}{2} \int^{t} dt' \left( \frac{9}{2}H + \frac{\dot{\omega}}{\omega} \right) \right] \no \\
&\times \exp \left[ \pm i\int^{t} dt'  m_{\rm eff} \right].
\end{align}
with $A_{\pm}$ constants. Using the fact that $H \equiv H_{\rm m}(t) = 2/(3t)$, and performing the integration in the first exponential we arrive at,
\be
\delta H_{\rm (osc)} \approx \sum_{\pm} \frac{ A_{\pm} }{ (\omega t^{3})^{1/2} } \exp \left[\pm i\int^{t} dt'  m_{\rm eff} \right]. \label{SolHomogPertMatterWKB}
\ee

The second integration can be performed after choosing a particular model. From (\ref{SolHomogPertMatterWKB}) one can see that the amplitude of the oscillating solution grows as one goes backwards in time, which is exactly the behavior pointed out for $f(R)$ models in Refs. \cite{Staro_Sing,Tsuj_f(R)Pert,Fro_Sing}, and was due to a curvature singularity as explained in Ref. \cite{Fro_Sing}. Therefore, $f(R,G)$ models suffer from the same problem too.

For the model giving a zero effective anisotropic stress during matter domination, see relations (\ref{eq:mattertoy}) and (\ref{eq:mattertoyExponents}), one can check that both $\omega$ and $m_{\rm eff}^2$ become complex, which is unphysical. What is more, the latter fact renders the analysis of homogeneous stability for this model impossible, or at least highly non trivial. One could possible seek to find a model without this peculiarity, however we shall leave this for future work.

We notice that the models of type (\ref{Zero_Anis_Stress_Models}) that have no anisotropic stress during a de Sitter phase with specific expansion rate $H_0=M/2$ will pass through $\omega=0$ and thus $m_{\rm eff} \rightarrow \infty$ in any background if the expansion rate $H(t)$ crosses this critical value $M/2$.
 
A different, general way to decrease the anisotropic stress is to move close to GR by decreasing the deviations from the extra $f(R,G)$ contributions, which effectively implies
\be
f_{RR}, f_{RG}, f_{GG}  \ll 1. \label{MatterConditions}
\ee
In this case, we also make $\omega$ small while $F\rightarrow 1$. Again this will lead to rapid oscillations, and we suspect that this is the reason for those seen in \cite{Pog_Pattern}. Once the genie of extra degrees of freedom is out of the bottle, it is difficult to push it back in without further complications.

\section{Toy models} \label{sec:ToyModels}
In this section we will study the de Sitter behavior for some characteristic cases of the class of models found in (\ref{Zero_Anis_Stress_Models}). For the sake of generality we will consider 
\be
\Omega = R + \epsilon \frac{G}{M^2}, 
\ee 
with $\epsilon = \pm 1$. The particular class of models with a vanishing of the anisotropic stress, found in (\ref{Zero_Anis_Stress_Models}), correspond to $\epsilon \rightarrow -1$ and $M \rightarrow 2 H_{0}$.

First note that, for the class of models (\ref{Zero_Anis_Stress_Models}), it is possible to parametrize both the de Sitter existence and stability conditions in terms of the parameter $\gamma$, which controls the different regimes of the model. For simplicity and illustration let us assume that $f_{2} = 0$. Then, the de Sitter condition (\ref{deSitterPointCondition}) becomes 
\be
 \left( \frac{\gamma + \epsilon}{2\gamma + \epsilon} \right)f_{\Omega} \Omega_{0} - f(\Omega_{0}) = 0, \label{dSpoint_Gamma}
\ee

with $f_{\Omega}\equiv f_{\Omega}(\Omega_{0})$, and
\be
\Omega_{0} = 6H_{0}^2 \left( \frac{2\gamma + \epsilon}{\gamma} \right). \label{Omega_Gamma}
\ee
Furthermore, for the de Sitter stability condition (\ref{dSStabilityCondInhomog}) we get
\be
\left( \frac{\gamma}{\gamma + \epsilon} \right)^{2} \frac{f_{\Omega}}{f_{\Omega \Omega}} \geq 2 \left( \frac{\gamma}{2 \gamma + \epsilon} \right) \Omega_{0}, \label{dSStability_Gamma}
\ee
We will assume that $\Omega_{0} > 0$, $\gamma > 0$ and real. The limits $\gamma \rightarrow \infty$ and $\gamma \rightarrow 0$ correspond to the pure $f(R)$ and $f(G)$ regimes respectively.

In principle, we will assume that through (\ref{dSpoint_Gamma}) we can express $\Omega_{0}$ in terms of $\gamma$ and the other possible parameters of the model as $\Omega_{0}=\Omega_{0}\left( \gamma, c_{i} \right)$, and then use (\ref{dSStability_Gamma}) to get a constraining condition.

\subsection{$ f(\Omega) = \Omega + \Omega \ln\left(\Omega/c^2 \right)  $ \label{sec:TMA}}
Here, $c$ is a positive constant of mass dimensions. This model is able to re-produce a late-time acceleration, since at late times $\Omega \ll 1$, and the logarithmic term will dominate. In four dimensions the linear term $\Omega$ is essentially equivalent to the Ricci scalar $R$ since the Gauss-Bonnet term does not contribute to the equations of motion. The absence of a Minkowski solution makes this model rather unrealistic.

A non trivial de Sitter solution can be found using (\ref{deSitterPointCondition}) 
\be
\Omega_{0} = c^{2} e^{\epsilon/\gamma},
\ee
and the Hubble parameter is then trivially given by (\ref{Omega_Gamma}).

The stability condition (\ref{dSStability_Gamma}) yields
\be
\frac{2\gamma^2 - 1}{2\gamma + \epsilon} \geq 0.
\ee
For both branches, $\epsilon = \pm 1$, de Sitter space is stable when $\gamma > \sqrt{2}/2$.

To illustrate the singularity when trying to reach de Sitter, we set $\gamma=1$ and for simplicity $c=\sqrt{6 e}$ so that the de Sitter solution is given by $H_0 = 1$. Expanding the equation of motion in $\delta H$ we find to first order,
\be
2 \left(1+2 (\delta \dot{H})^2  \right) \delta H + O\left( (\delta H)^2 \right) = 0 .
\ee
Only in the second order term a contribution $\delta \ddot{H} (\delta H)^2$ appears. We notice that it is not possible to solve the first term for real $\delta H$, so that necessarily $\delta \ddot{H} \propto 1/\delta H$ will diverge when we try to dynamically reach de Sitter. The only exception is $\delta H = 0$, i.e. the solution that is always de Sitter. 

If we are looking at a model that is close to the critical case but has not quite zero anisotropic stress, $\gamma = 1+\varepsilon$, then the stability analysis indicates that model is stable as long as $\varepsilon>-0.29$, and indeed numerically we observe rapid oscillations of $\delta H$ around de Sitter for most cases. For very small $\varepsilon$ we can get an idea of the model behavior by first solving the equation of motion for $\delta\ddot{H}$ and then linearizing the full equation with respect to $\delta H$, under the assumption that we will look at the evolution close to de Sitter. We can then expand the resulting equation in $\varepsilon$, which to lowest order in $\varepsilon$ and for $c=\sqrt{6 e}$ results in
\be
\delta\ddot{H} \approx \frac{2}{\varepsilon^2} \delta H \left(2 (\delta\dot{H})^2-1 \right) .
\ee
We see that the second time derivative of $\delta H$ will become very large as $\varepsilon\rightarrow 0$, which shows again the presence of the singularity for the critical case of zero anisotropic stress. But we also see that even close to de Sitter the evolution of $\delta H$ is governed by a non-linear differential equation since the time derivatives of $\delta H$ will in general be large for models with small anisotropic stress and so cannot be neglected. A detailed study of how small the anisotropic stress can be made is therefore not straightforward and left to future work.

\subsection{$f(\Omega) = \Omega + c\Omega^{n}$}
In the context of $f(R)$ gravity, models of this type were suggested as an explanation for late time acceleration \cite{Capo_f(R)Model,Car_f(R)Model} with $n<0$, while models with $n>0$ can lead to acceleration at early times and explain inflation. Furthermore, it was found that de Sitter space is unstable unless $cn<0$ \cite{Faraoni_Stabilityf(R)}. Here, we assume that $cn>0$, otherwise the no--ghost condition $F>0$ could be violated. 

The de Sitter point equation (\ref{dSpoint_Gamma}) gives two solutions, namely $\Omega_{0} = 0$ (for $n>0$) which corresponds to Minkowski spacetime and a non trivial de Sitter one,
\be
\Omega_{0} = \left[ \frac{\gamma}{c(\gamma(n-2) + \epsilon(n-1))} \right]^{1/(n-1)}.
\ee
In order for $\Omega_{0}$ to be real and positive one has to ensure that the quantity in the denominator in the latter relation is positive. We shall also require that the Hubble parameter, as given implicitly in relation (\ref{Omega_Gamma}), will be real and positive too. 

For both branches $\epsilon = \pm 1$, de Sitter is always unstable when $n<0$. For $n > 0$, it is always unstable if $\epsilon = 1$, but for $\epsilon = -1$, $n > 2$, the stability condition (\ref{dSStability_Gamma}) gives 
\be
\frac{n-1}{n-2} < \gamma < \frac{n + \sqrt{n/2} - 1}{n-2},
\ee
with both $\Omega$ and $H_{0}$ being real and positive.

To avoid a superluminal sound speed, the model should lie in the $f(R)$ regime, characterized by $\gamma > 1$, which is satisfied here, as the right branch of above inequality approaches the value $1^{+}$ as $n \rightarrow \infty$. Further, for $n>2$, Minkowski space is always stable.

To consider the equation of motion close to de Sitter, we set $\epsilon = -1$, $\gamma =1$ and choose $c = - 6^{(1-n)}$, for which $H_0=1$. We also assume that $n \neq 1$. We again expand in $\delta H$. The lowest order equation becomes now
\be
\left(1+ 2 n (\delta \dot{H})^2  \right) \delta H + O\left( (\delta H)^2 \right) = 0 .
\ee
Again the second derivative of $\delta H$ appears only at order $(\delta H)^2$. This time we can in principle make the first order term vanish for $n<0$, which would allow to cross $\delta H=0$ with a finite second derivative. However, there are two problems: Firstly, we can only cross, not move into and stay on $\delta H=0$, since locally we need $\delta H \sim (t-t_0)/\sqrt{- 2 n}$ to avoid triggering the instability, and secondly this requires an infinite amount of fine-tuning in the initial conditions: we need to reach de Sitter at exactly the right speed, else we are either repelled, or a catastrophe engulfs the universe. So in reality again it is impossible to reach de Sitter dynamically.

\subsection{ $f(\Omega) = \Omega + c_{0} \lambda \left( (1+\frac{\Omega^2}{c_{0}^2})^{-n} - 1 \right) $}
This is a straightforward generalization of Starobinsky's disappearing cosmological constant model \cite{Staro_Sing}. It was proposed in the context of $f(R)$ gravity as a late time acceleration model, that has a vanishing cosmological constant in Minkowski spacetime. It is trivial to check that Minkowski, $f(0) = 0$, is indeed a solution, but unstable since $f_{\Omega \Omega}(0)<0$.

The model is characterized by three parameters, $c_0 $, $\lambda$, $\gamma>0$.

From the de Sitter point equation, one can find an expression for $\lambda$ as a function of $\gamma$, and $x_{1}\equiv \Omega_{0}/c_{0}$
\be
\lambda = \frac{x_{1}(g_{0} - 1)(1+x_{1}^2)^{n+1}}{\left[x_{1}^2(2ng_{0}+1)-(1+x_{1}^2)^{n+1} + 1 \right]}, \label{lambda_Starob}
\ee
where $g_{0} \equiv (\gamma+\epsilon)/(2\gamma+ \epsilon)$. Taking the limit $\gamma \rightarrow \infty$ in the above expression one recovers the one given in Starobinsky's paper Ref. \cite{Staro_Sing}. 

Let us assume that $c_0$ is of the order of the de Sitter scale, $\Omega_{0}/c_0 \equiv x_{1} = 1$. The de Sitter stability condition then reads
\be
2(\gamma + \epsilon)n^2 + (2\gamma + 1)n + (2\gamma + \epsilon)(1-2^n) \leq 0,
\ee
For $n = 1$ de Sitter is stable if
\be
-\frac{2\gamma}{2\gamma + \epsilon} \geq 0,
\ee
which is never satisfied for both branches $\epsilon = \pm 1$. However, choosing $x_{1} = 1/2, n = 1$, we find that de Sitter is stable for $\epsilon = 1$ and $\gamma > 1/2$, as well as for $\epsilon = -1$ and $\gamma \gtrsim 0.68$.

Stability can be established for a wide range of the model parameters, but that would require a detailed exploration of the parameter space of $\{ c_{0}, \lambda, n\}$, and we are not interested in this here.

For the critical case $\epsilon=-1$, $\gamma=1$, choosing the $\lambda$ of (\ref{lambda_Starob}) and in addition $c_0 = 6 H_0^2$ for simplicity, we find to first order in $\delta H$
\be
\left[ 1-\frac{n\left(H_0^4 - 2 n (\delta \dot{H})^2 \right)}{(2^n-1) H_0^4} \right] \delta H + O\left( (\delta H)^2 \right) = 0 .
\ee
This equation is of the same kind as the one found for the previous toy model, and it leads to the same behaviour. The special case $n=1$ leads to  the equation $(\delta \dot{H})^2 \delta H = 0$, which prohibits any crossing of $\delta H =0$ as otherwise $\delta \ddot{H}$ has to diverge.


\section{Conclusions}

In this paper we study the anisotropic stress in $f(R,G)$ type modified gravity models. We investigated the possibility of finding models that are able to mimic GR at least in the sense that they do not create an additional, effective contribution to the anisotropic stress, i.e $\phi = \psi$ in the Newtonian gauge. For the needs of our analysis, we also derived the necessary background stability conditions. We started by considering the case of a de Sitter background, since this allowed us to find the general class of models with vanishing anisotropic stress. The de Sitter case is in addition interesting as current observations indicate that the Universe is approaching this state. We further considered the general case in the small-scale limit, and in more detail the case of a matter dominated expansion.

We find that for de Sitter expansion, the anisotropic stress is inextricably linked to the presence of an extra scalar degree of freedom. Suppressing the effective, geometric anisotropic stress is equivalent to suppressing the extra degree of freedom, which either requires the model to revert back to GR or else leads to an instability in the background evolution. In addition, it leads to problematic effects like rapid oscillations of the gravitational potential and the curvature perturbation (with possible runaway production of scalar particles) and superluminal propagation of the perturbations. The same problems appear when one tries to generally decrease the extra degrees of freedom through a model reconstruction, in order to obtain an evolution similar to GR. We think that this has been observed for numerically reconstructed $f(R)$ models in a matter dominated background \cite{Pog_Pattern}, indicating that it is more general and not restricted to de Sitter.

Furthermore, our stability analysis reveals that the curvature singularity present in $f(R)$ models \cite{Staro_Sing,Tsuj_f(R)Pert,Fro_Sing} appears in the more general $f(R,G)$ case as well. What is more, its  unwanted effect on the behavior of curvature perturbation is amplified for all models that try to suppress the anisotropic stress by decreasing $f_{RR}$, $f_{RG}$ and $f_{GG}$. In these cases we find rapid curvature oscillations with arbitrarily high amplitude as $\phi-\psi \rightarrow 0$.

In the case of a pure matter dominated background, we were able to construct an explicit model that gives a zero effective anisotropic stress in the subhorizon limit. At late times, when the gravity modifications are expected to appear and the evolution ceases to be matter dominated, this model will no longer give $\phi = \psi$. This could possibly be avoided by constructing such models for a whole expansion history including late-time accelerated expansion. However, such a procedure would necessarily involve significant fine-tuning as changes in the expansion rate would have to coincide with changes in the behavior of the function $f(R,G)$, which would in general depend sensitively on initial conditions. This appears to be rather difficult to construct. In addition, as discussed above, such a model would not be able to reach the de Sitter state without encountering a singularity.

While the link between effective anisotropic stress and the scalar degree of freedom of the theory was studied here in the context of $f(R,G)$ models, it is also present in scalar-tensor and DGP models: If a scalar-tensor model is coupled to the Ricci scalar in the action through $F(\varphi) R$ then the anisotropic stress is proportional to $(F'/F) \delta\varphi$ and the analogy to the $f(R)$ case is obvious. In DGP, the effective anisotropic stress vanishes for $r_c \propto M_4^2/M_5^3 \rightarrow\infty$ where $M_4$ and $M_5$ are the four- and five-dimensional Planck scales \cite{koyma,luestark}. In this limit, the 5-dimensional part of the action is suppressed and only the usual 4D Einstein-Hilbert action remains.

We conjecture that suppressing the effective anisotropic stress in modified gravity models is difficult, if not impossible, to achieve in a realistic scenario. In models with a single extra degree of freedom that we looked at ($f(R)$, $f(G)$, scalar-tensor models and DGP) it is not possible at all to have no effective anisotropic stress except in the GR limit. In more complicated cases like $f(R,G)$ it is possible to cancel the contributions to the effective anisotropic stress coming from several extra degrees of freedom, but this appears to be fine tuned and the resulting models tend to develop fatal singularities. This reinforces the role of the anisotropic stress as a key observable for current and future dark energy surveys. While the observation of a strong anisotropic stress would point towards a modification of GR, the absence of anisotropic stress would present a significant challenge for modified gravity models and would require strong fine-tuning, which in turn favors scenarios where the dark energy is a cosmological constant or an extra minimally-coupled field with negative pressure.

\begin{acknowledgments}
It is a pleasure to thank Luca Amendola, Mark Hindmarsh and Andrew Liddle for stimulating discussions. MK acknowledges financial support from the Swiss NSF. IDS is supported by GTA funding from the University of Sussex. IDS would also like to thank the University of Geneva for the warm hospitality, while part of this work was being prepared.

\end{acknowledgments}

\appendix

\section{Homogeneous perturbations of $f(R,G)$} \label{sec:HomogPert}
In this section we will present the stability analysis of any fixed point of the the $f(R,G)$ Friedmann equation, using homogeneous perturbations
around the relevant solution.
Our starting point is the $t-t$ equation (\ref{f(R,G)FriedmanEquation}), which for convenience we reproduce it here again,
\be
3H^2F+3H\dot{F}+12H^{3}\dot{\xi}-\frac{1}{2}V - \rho_{i}= 0. 
\ee

If $H \equiv H(t)$ is a solution of above equation then perturbing around it as $H(t) \rightarrow H(t) + \delta H(t)$, and keeping up to first order terms
we get for the curvature scalars and their first time derivatives respectively
\be
 R \rightarrow R + 6\left( 4 H \delta H + \delta \dot{H}  \right),
 \ee
 \be
 G \rightarrow G+ 24\left[ 2(2H^3 + \Ht H)\delta H + H^2 \delta \dot{H} \right].
 \ee
The next step is to perturb the modified Friedman equation (\ref{f(R,G)FriedmanEquation}). Particularly, the scalar potential becomes
\begin{align}
V & \rightarrow V + R \delta{F} + G \delta{\xi} \nonumber \\ 
&=V + ( R F_{R}+G\xi_{R} )\delambda + (R F_{G}+G\xi_{G})\desigma \nonumber \\ 
&\equiv V +V_{(R)}\delambda+V_{(G)}\desigma,  
\end{align}
and after evaluating the scalar field perturbations, it takes the form
\begin{align}
 V \rightarrow \; & V + 6 \left( V_{(R)} + 4H^2 V_{(G)} \right) \dedH + 24 \big( \, H V_{(R)}  \no \\
 & + 2(2H^3 + \dot{H}H ) V_{(G)} \, \big) \deH,
\end{align}
where subscripts in brackets simply denote indices, while those outside brackets denote derivative with respect to the corresponding variable.

Using relations given above, and after some algebra, the modified Friedman equation becomes
\be
C_{(H)} \delta H + C_{(R)}\delta R  + C_{(G)} \delta G + C_{\dot{(R)} }\delta \dot{R} + C_{\dot{(G)}} \delta \dot{ G } - \delta \rho_{i} = 0,
\ee
with 
\begin{align}
& C_{(H)} \equiv  6H \Fn + 3\Fnt + 36 H^2 \xint ,\\
& C_{(R)} \equiv 3H^2 \Fln + 3H \Flnt + 12 H^3 \xilnt - \frac{1}{2} V_{(R)} ,\\
& C_{(G)} \equiv 3 H^2 \Fsn + 3H\Fsnt + 12H^3 \xisnt - \frac{1}{2} V_{(G)} , \\
& C_{\dot{(R)}} \equiv 3 H \Fln + 12 H^3 \xiln ,\\ 
& C_{\dot{(G)}} \equiv 3 H \Fsn + 12 H^3 \xisn.
\end{align}
Substituting for the perturbations of $\delta R$, $\delta G$ and their derivatives we arrive at
\be
C_{1} \delta \ddot{H}  + C_{2} \delta \dot{H} + C_{3} \delta H - \delta \rho_{i} =0, \label{HomogPertGeneral}
\ee
with
\begin{align}
& C_{1} \equiv 6 \left[ C_{\dot{(R)}} + 4C_{\dot{(G)}} H^2   \right], \label{C1general}\\
& C_{2} \equiv 6 \left[ C_{(R)} + 4C_{(G)} H^2 + 4 C_{\dot{(R)}} H + 16 C_{\dot{(G)}} (H^3 + \Ht H)  \right],\label{C2general}\\
& C_{3} \equiv 6 \Big[ H \Fn + \frac{1}{2} \Fnt + 6 H^2 \xint + 4 C_{(R)} H + 4C_{(\dot{R})}\dot{H}  \no \\ 
& \; \; \; \; \; \; + 8C_{(G)}(2 H^3 + \Ht H) + 8C_{\dot{(G)}} (6 H^2 \Ht + \Ht^2 + \Htt H) \Big]. \label{C3general}
\end{align}

Defining $\omega \equiv F_R + 8 H^2 ( F_G + 2 H^2 \xi_G)$, the generalisation of Eq.~(\ref{omega}) for arbitrary $H$, we find that always
\be
C_1 = 18 H \omega .
\ee
For a polynomial background expansion, described by $a(t) \propto t^{p}$, the other coefficients become
\hspace{-1cm}\begin{align}
&C_{2} = \frac{18H}{p} \left[ p\dot{\omega} + 8H^3(1+3p)(\xi_{R} + 2H^2 \xi_{G}) + (1+3p)HF_{R} \right] , \\
&C_{3} = 3 \Big\{  \dot{F} + 2HF  + 12 H^{2} \Big[  \frac{4(5p^{2}+2)}{p^3} H^3(\xi_{R} + 4H^2\xi_{G})   \no \\
&- \frac{4}{p}H^2(\dot{\xi}_{R}+ 4H^2 \dot{\xi}_{G}) - 2(H \omega - \dot{\omega}) + \dot{\xi}  \Big] \Big\}.
\end{align}
For a de Sitter expansion, $a(t) \propto \exp[H_{0}t]$, and $H=H_{0} = const.$, we get
\begin{align}
&C_{2} = 3H_{0}C_1\\
&C_{3} = \left( \frac{F}{3 \omega} - 4H_{0}^2 \right)  C_1.
\end{align}

\section{Inhomogeneous perturbations and de Sitter stability}\label{sec:InhomPertApp}
The general metric element for scalar perturbations around a flat FRW background reads
\begin{align}
ds^2 = & -(1+2\alpha)dt^2 - 2a(t)\partial_{i}\beta \, dt\, dx^{i} \no \\
& +a(t)^{2}\left( \delta_{ij} -  2\phi\delta_{ij} + 2\partial_{i}\partial_{j} \gamma  \right)dx^{i}dx^{j}. \label{PertMetricGen}
\end{align}

The general form of scalar perturbation equations around FRW for $f(R,G)$ models can be found in Ref. \cite{FeliceSuyama}. Here, we shall present the full set of equations for the case of de Sitter space only. 

Before we proceed, let us define the gauge invariant variable $\Phi$ as
\be
\Phi \equiv \Phi(t) \equiv \frac{\delta F + 4H^2 \delta \xi}{2F}, \label{Phi_GI}
\ee
with $H \equiv H_{0}$ as well as the rest of the background quantities evaluated on the de Sitter point. The perturbation equations then read as
\be
 3H^2 \psi + \ka \left( H \chi + \phi \right)  +  3H\dot{\phi}  =  3H \dot{\Phi} + (\ka - 3H^2)\Phi , \label{Inhom_Pert_00}
 \ee
 \be
 H \psi +   \dot{\phi}  = \dot{\Phi} - H \Phi  , \label{Inhom_Pert_0i}
  \ee
 \be
 \dot{\chi} + H \chi + \phi - \psi  = 2 \Phi , \label{Inhom_Pert_ij}
\ee
\begin{align}
\delta R = -2 \Big[ &12H^2 \psi + 3\ddot{\phi} + 12H \dot{\phi} + 3H\dot{\psi} \no  \\
&   \ka \left(\dot{\chi} + 2H\chi + 2\phi - \psi \right) \Big] , \label{R_Variation}
\end{align}
\begin{align}
\delta G = -8 \Big[& 12H^4\alpha - 3H^2 \ddot{\phi} + 3H^3 \dot{\alpha} - 12H^3\dot{\phi} \no \\
 &+ \ka H^2 \left(2H \chi + \chi- 2\phi - \alpha \right) \Big].
\end{align}
Equations (\ref{Inhom_Pert_00}), (\ref{Inhom_Pert_0i}) and (\ref{Inhom_Pert_ij}) correspond to the $00$, the $0i$ and the $ij (i \neq j)$ components respectively. Particularly, equation (\ref{Inhom_Pert_ij}) is the anisotropy equation, and the choice of variable $\Phi$ is now evident: it is the r.h.s of the latter equation, describing the effective anisotropic stress in de Sitter space, $\Phi = \Pi^{{\rm (eff)}}$, and therefore is gauge invariant. 

In order to re-express above equations in terms of gauge invariant variables only, we need a second gauge invariant variable apart from $\Phi$. Following \cite{FeliceSuyama} we define
\be
\Psi \equiv \Phi + \phi - H \chi. \label{Psi_GI}
\ee 
Now, using equation (\ref{Inhom_Pert_0i}) in (\ref{Inhom_Pert_00}) we get
\be
\Phi = \phi + H\chi, \label{Phi_GI2}
\ee
which can be inserted into (\ref{Psi_GI}) to give
\be
\Psi = 0.
\ee 
Using equations (\ref{Inhom_Pert_0i}), (\ref{Inhom_Pert_ij}) as well as (\ref{Phi_GI2}) we can re-express the curvature perturbation in terms of the gauge invariant potential $\Phi$
\be
\delta R = -6 \left[ \ddot{\Phi} + 3H\dot{\Phi} + \left( \frac{k^2}{a^2} - 4H_{0}^2 \right) \right]. \label{GI_deltaR}
\ee

\section{Sub-horizon solution for $\Phi$ in the WKB approximation\label{app:wkb}}
Considering the evolution equation (\ref{MasterEquation}) in de Sitter space for the gauge invariant potential $\Phi$, we assume a solution of the form
\be
\Phi(t) = C\rm{e}^{i \theta(t)},
\ee
with $C$ a constant, and $\ddot{\theta}(t) \ll 1$. Then, we can calculate that
\bea
\Phi(t) & \approx&  \sum_{\pm} C_{\pm} \exp \left[ i \int^{t} dt' \dot{\theta}^{\pm}(t') \right]  \label{WKBdeSitterSolution}  \\
& \equiv & \sum_{\pm} C_{\pm} \exp\left\{  \int^{t} dt' \left[ -\left(A + \frac{\dot{B}(t')}{B(t')} \right) \pm i B(t')   \right]  \right\} \nonumber
\eea
with
\bea
A \equiv \frac{3H_{0}}{2},  \; \; B(t) \equiv \frac{1}{2} \sqrt{4\left(k^2e^{-2H_{0}t} + m^{2}_{\rm eff} \right)  -  9H_{0}^2}
\eea

From solution (\ref{WKBdeSitterSolution}) we can calculate the limit when $m^2_{\rm eff} \gg 1$, which is the case when $\Pi^{\rm (eff)} \rightarrow 0$. In this case we have,
\be
B(t) \approx m_{\rm eff}, \; \; \dot{B}(t) \approx 0,
\ee
and the solution is approximately given by
\be
\Phi(t) \approx \sum_{\pm} C_{\pm} \exp \left[ - \frac{3}{2}H_{0}t \; \pm \;  i m_{\rm eff} t \right]. 
\ee

\end{document}